\documentclass[aps, prl, reprint, groupedaddress, amsmath, amssymb]{revtex4-2}

\usepackage{graphicx}
\usepackage{dcolumn}
\usepackage{bm}
\usepackage{comment}
\usepackage{xr}
\begin{document}



\title{Polarity-selective domain wall propagation driven by spin-orbit torque}


\author{Nobuyuki Umetsu}
\affiliation{Frontier Technology R\&D Institute, Kioxia Corporation, 3-13-1, Moriya-cho, Kanagawa-ku, Yokohama, 221-0022, Japan}
\author{Michael Quinsat}
\affiliation{Frontier Technology R\&D Institute, Kioxia Corporation, 3-13-1, Moriya-cho, Kanagawa-ku, Yokohama, 221-0022, Japan}
\author{Masaki Kado}
\affiliation{Frontier Technology R\&D Institute, Kioxia Corporation, 3-13-1, Moriya-cho, Kanagawa-ku, Yokohama, 221-0022, Japan}
\author{Hiroki Tokuhira}
\affiliation{Frontier Technology R\&D Institute, Kioxia Corporation, 3-13-1, Moriya-cho, Kanagawa-ku, Yokohama, 221-0022, Japan}

\date{\today}

\begin{abstract}
Spatial variations in magnetic anisotropy impede the
spin-orbit-torque-driven motion of chiral domain walls
in perpendicularly magnetized nanowires.
We show that the local geometry of a two-dimensional
anisotropy landscape can select which wall polarity
propagates under a given current.
Micromagnetic simulations of spatially correlated random
anisotropy landscapes generated by diffusion-based
smoothing reveal different propagation outcomes for
up--down and down--up walls within the same disorder
realization.
A single inclined anisotropy boundary isolates the
mechanism: in the presence of the Dzyaloshinskii--Moriya
interaction, opposite wall polarities develop opposite
current-induced tilts, producing different alignment
with the boundary and hence different pinning strengths.
Reversing the boundary inclination interchanges the pinned
and propagating polarities.
An extended collective-coordinate model captures this
geometric selection and distinguishes the propagation
threshold from the static pinning limit.
\end{abstract}

\maketitle

Spatial inhomogeneity in magnetic properties plays a central role in the dynamics of magnetic textures in thin films.
Magnetic domain walls (DWs) can be trapped in local minima of the associated energy landscape,
and such pinning affects their dynamics over a wide range of driving conditions
\cite{Fukami_2013, Jeudy_2018, Jagt_2022, Yoon_2022, Giuliano_2023, Balan_2023, Albornoz_2024, Joseph_2026, Bhattacharjee_2026, Karim_2026}.
Understanding these effects is essential for the precise and energy-efficient control of DW motion in spintronic devices
\cite{
Parkin_2015, Kumar_2022, Venkat_2023, Umetsu_2025_PRB, Quinsat_2026, Jeon_2026}.

Current-induced DW tilting provides a possible coupling
between spin-orbit torque (SOT)-driven motion and spatial inhomogeneity.
In heavy-metal (HM)/ferromagnet (FM) heterostructures with
perpendicular magnetic anisotropy (PMA), a charge current
in the HM generates a transverse spin current through the
spin Hall effect, exerting SOT on the FM magnetization.
The interfacial Dzyaloshinskii--Moriya interaction (DMI) stabilizes chiral N\'{e}el DWs \cite{
Dzyaloshinsky_1958, Moriya_1960, Thiaville_2012, Khvalkovskiy_2013},
enabling efficient current-driven DW motion via SOT \cite{Moore_2008, Kim_2010, Miron_2011, Haazen_2013, Emori_2013, Ryu_2013, Torrejon_2014, Ryu_2014, Yang_2015, Torrejon_2016, Krishnia_2017, Guan_2021, Wang_2022, Barker_2023, Umetsu_2025_JMMM}.
Such DWs are known to exhibit current-induced in-plane tilting \cite{
Ryu_2012, Boulle_2013, Emori_2014, Martinez_2014, Conte_2017, Baumgartner_2018, Kim_2022}.
By changing the DW orientation relative to the underlying
inhomogeneity, this tilting can modify the pinning strength.

Previous models have often restricted magnetic-property
variations to the nanowire axis
\cite{Franken_2011, Zhang_2017, Moretti_2017,
Giuliano_2023, Joseph_2026}.
Such descriptions miss variations across the width, which
produce nonuniform pinning along a tilted DW.
Capturing this coupling requires a two-dimensional
anisotropy landscape and a description of both DW
translation and tilting.

Here, we investigate SOT-driven DW motion using
micromagnetic simulations with two-dimensional,
spatially correlated random PMA landscapes generated
by diffusion-based smoothing of independent Gaussian
cell noise.
To identify the origin of polarity-dependent propagation
in these landscapes, we consider a single inclined
PMA boundary and analyze the DW dynamics using an
extended one-dimensional model (1DM) with three
collective coordinates: the DW position, magnetization
angle, and tilt angle
\cite{Boulle_2013, Martinez_2014, Volkov_2021}.
We show that opposite current-induced tilts produce
different alignment with the boundary and hence
different pinning strengths for the two DW polarities.

We consider an ultrathin FM nanowire extending along $x$,
with width $L_y$ and thickness $L_z$, on an HM layer.
The FM exhibits PMA, and DMI arises at the HM/FM interface.
A charge current applied along $x$ exerts spin torques on the magnetization.
The unit magnetization $\bm{m}$ obeys the Landau--Lifshitz--Gilbert equation,
\begin{equation}
\dot{\bm{m}}
= - \frac{\gamma}{M_\mathrm{s}}\frac{\delta \mathcal{E}}{\delta \bm{m}}   \times \bm{m} 
+ \alpha \bm{m}\times \dot{\bm{m}}
+ \bm{\tau}_{\rm s},
\end{equation}
where $\gamma$ is the gyromagnetic ratio,
$M_{\rm s}$ is the saturation magnetization,
$\mathcal E$ is the magnetic energy density,
$\delta\mathcal E/\delta\bm m$ denotes its functional
derivative with respect to $\bm m$,
$\alpha$ is the Gilbert damping constant, and
$\bm\tau_{\rm s}$ is the current-driven spin torque.
The magnetic energy includes demagnetization, exchange, PMA,
and DMI contributions. The latter three are characterized by
the exchange stiffness $A$, PMA constant $K_{\rm u}$, and DMI
constant $D$, respectively.
The spin torque $\bm{\tau}_{\rm s}$ includes spin-transfer
torque (STT) and the damping-like and field-like components
of SOT \cite{Risinggard_2017, Umetsu_2025_JMMM,
Umetsu_2025_PRB}:
\begin{equation}
\begin{aligned}
\bm{\tau}_{\rm s}
={}&u\frac{\partial\bm m}{\partial x}
+\gamma\mu_0\xi_{\rm so}j_{\rm HM}
\Bigl[
\bm m\times(\bm m\times\bm e_y)\\
&\hspace{6em}
-\beta_{\rm so}\bm m\times\bm e_y
\Bigr].
\end{aligned}
\end{equation}
Here, $u=\mu_{\rm B}Pj_{\rm FM}/(M_{\rm s}e)$ is the
spin drift velocity, and
$\xi_{\rm so}=\hbar\Theta_{\rm SH}/(2\mu_0M_{\rm s}L_ze)$
is the damping-like SOT efficiency.
The parameter $\beta_{\rm so}$ is the signed ratio of
the field-like to damping-like effective-field amplitudes.
The current densities in the FM and HM are $j_{\rm FM}$
and $j_{\rm HM}$, respectively, and $\bm e_y$ is the unit
vector along $y$.
The remaining constants are the Bohr magneton $\mu_{\rm B}$,
spin polarization $P$, vacuum permeability $\mu_0$,
reduced Planck constant $\hbar$, spin Hall angle
$\Theta_{\rm SH}$, and elementary charge $e$.

The field-like torque is not the primary driving torque
and is reported to be smaller than the damping-like
torque in the relevant material system \cite{Lau_2017}.
To focus on damping-like SOT-driven propagation, we set
$\beta_{\rm so}=0$ for the results presented in the main text.
The effect of field-like torque on the propagation threshold and
polarity splitting is examined in
Sec.~\ref{sec:FLT} of the Supplemental Material.

We consider W/CoFeB/MgO, in which SOT-driven DW motion has been experimentally observed \cite{Umetsu_2025_JMMM, Torrejon_2014, Torrejon_2016}.
Unless otherwise stated, the parameters are
$M_{\rm s}=1.1\,\mathrm{MA/m}$,
$A=15\,\mathrm{pJ/m}$,
$D=0.34\,\mathrm{mJ/m^2}$,
$\alpha=0.03$,
$\Theta_{\rm SH}=-0.2$, and $P=0.5$.
We assume equal resistivities for the FM and HM, so that $j_{{\rm FM}}=j_{{\rm HM}}$ \cite{Umetsu_2025_JMMM}.
Hereafter, we write $j=j_{\rm FM}=j_{\rm HM}$.

\begin{figure}[htbp]
\centering{}\includegraphics[width=1\linewidth]{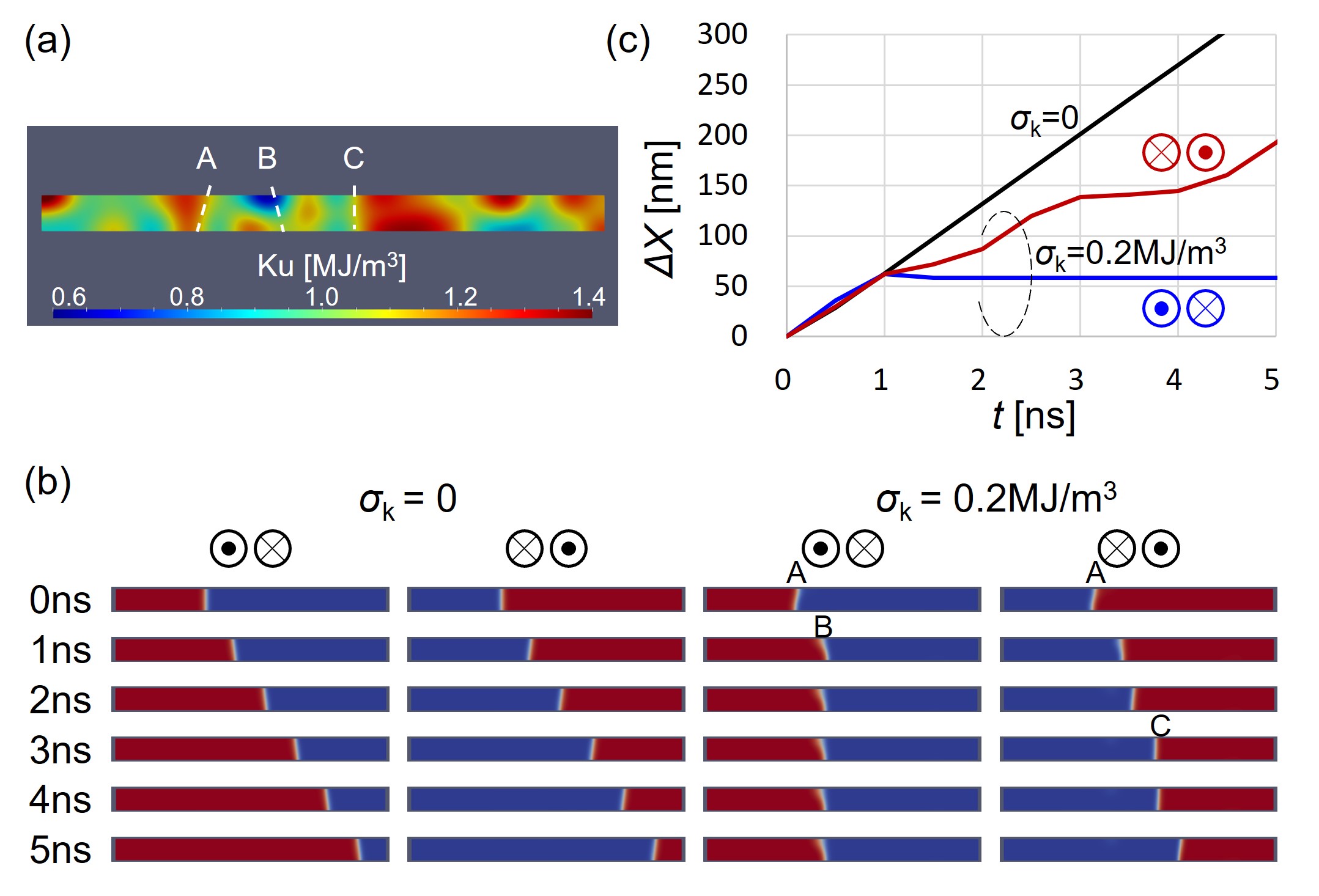}
\caption{
(a) Random PMA landscape with
$\bar K=1.06\,\mathrm{MJ/m^3}$,
$\sigma_{\rm k}=0.2\,\mathrm{MJ/m^3}$,
and $l_{\rm k}=20\,\mathrm{nm}$.
A is the relaxed initial DW position, and B and C are pinning sites.
(b) DW snapshots at $j=45\times10^{10}\,\mathrm{A/m^2}$ for the $\odot\otimes$ and $\otimes\odot$ walls
in uniform and disordered cases.
(c) Time evolution of DW displacement.}
\label{fig:LLG_dynamics-1}
\end{figure}

We perform micromagnetic simulations in an FM nanowire
with $L_x=600\,\mathrm{nm}$, $L_y=50\,\mathrm{nm}$ and $L_z=0.8\,\mathrm{nm}$.
Random PMA landscapes are generated by diffusive smoothing
of independent Gaussian cell noise, as described in
Sec.~\ref{sec:PMA} of the Supplemental Material.
The landscapes are parameterized by the mean $\bar K$,
nominal fluctuation amplitude $\sigma_{\rm k}$, and correlation
length parameter $l_{\rm k}$.
Figure~\ref{fig:LLG_dynamics-1}(a) shows a representative map of $K_{{\rm u}}$
for $\bar{K}=1.06 \,\mathrm{MJ/m^3}$,
$\sigma_{\rm k}=0.2\,\mathrm{MJ/m^3}$, and $l_\mathrm{k}=20\,$nm.
In the disordered case, the DW is initially  positioned at the local energy minimum labeled A in Fig.~\ref{fig:LLG_dynamics-1}(a).

Figure~\ref{fig:LLG_dynamics-1}(b) compares current-driven DW motion at $j=45\times10^{10}\,\mathrm{A/m^2}$
for up--down ($\odot\otimes$) and down--up ($\otimes\odot$) DWs in the uniform system
($K_{\rm u}=\bar K$, $\sigma_{\rm k}=0$) and
the disordered system shown in Fig.~\ref{fig:LLG_dynamics-1}(a).
The corresponding time evolution of the DW displacement, extracted from the spatial average of $m_z$ over the entire system, is shown in Fig.~\ref{fig:LLG_dynamics-1}(c).

\begin{figure}[htbp]
\centering{}\includegraphics[width=1\linewidth]{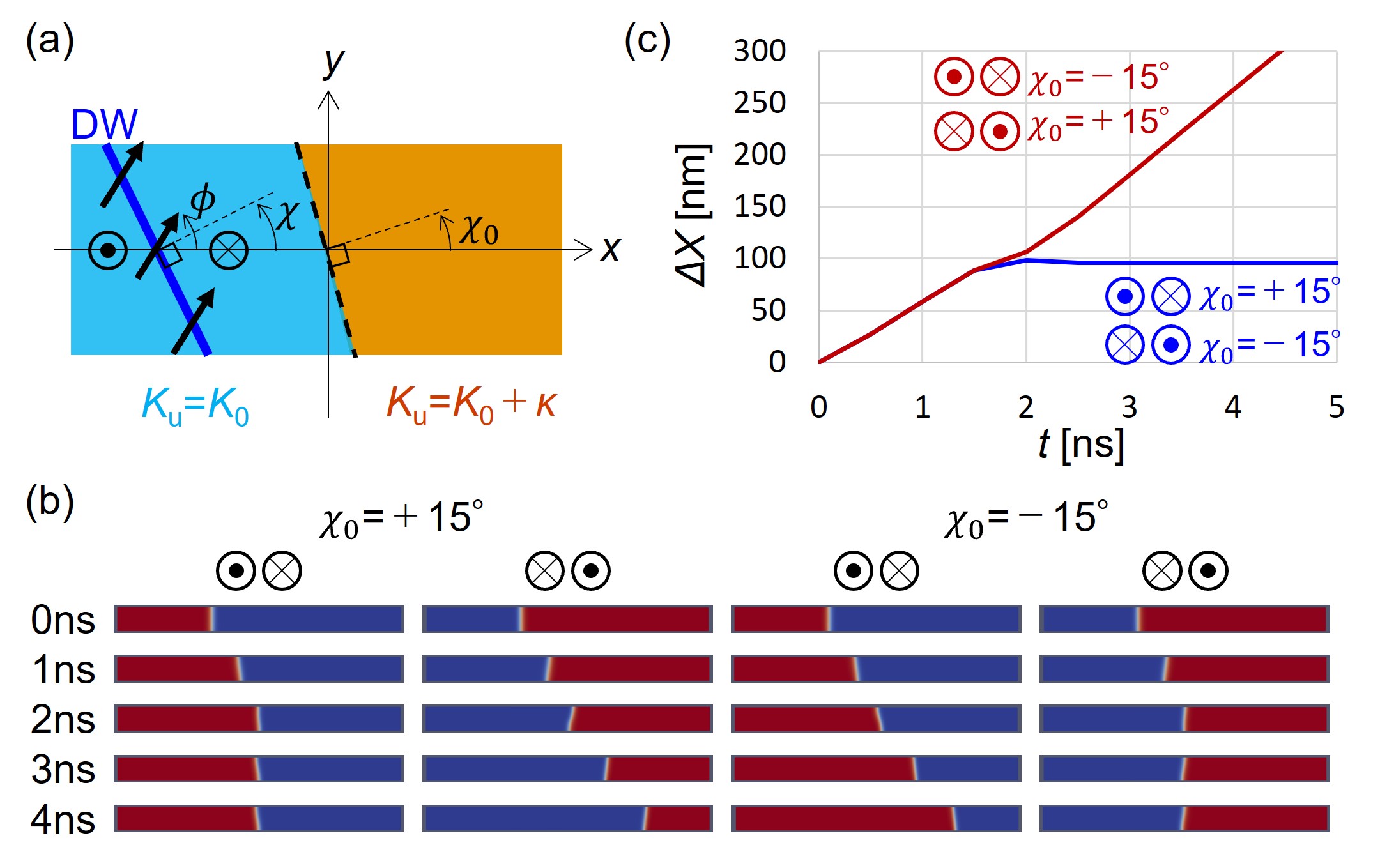}\caption{
(a) Step-like PMA profile.
(b) DW snapshots at $j=70\times10^{10}\,\mathrm{A/m^2}$ for $K_{0}=1.06\,\mathrm{MJ/m^3}$,
$\kappa=0.25\,\mathrm{MJ/m^3}$ and $\chi_{0}=\pm 15^{\circ}$.
(c) Time evolution of DW displacement.}
\label{fig:LLG_dynamics-2}
\end{figure}

We define the DW tilt angle $\chi$ as the signed angle
of the wall normal relative to the $x$ axis, with
positive angles measured counterclockwise in the
$x$--$y$ plane [Fig.~\ref{fig:LLG_dynamics-2}(a)].
In the uniform system, the two polarities have identical
velocities but opposite tilt angles: $\chi>0$ for the
$\odot\otimes$ wall and $\chi<0$ for the
$\otimes\odot$ wall.
The tilt direction depends on the signs of the DMI and
spin Hall angle \cite{Martinez_2014, Kim_2022}.

By contrast, in the disordered system ($\sigma_{\rm k}=0.2\,\mathrm{MJ/m^3}$),
the DW velocity depends on DW polarity.
As shown in Fig.~\ref{fig:LLG_dynamics-1}(b),
the $\odot\otimes$ wall becomes pinned at B,
whereas the $\otimes\odot$ wall slows transiently at B,
crosses it, and slows again at C.
Both B and C correspond to pinning sites associated with sharp spatial variations in $K_{{\rm u}}$.

We perform simulations for both DW polarities over 50 random realizations of $K_{{\rm u}}$.
Histograms of the DW displacement after a $10\,\mathrm{ns}$ current pulse
are shown in Fig.~\ref{fig:histogram} of the Supplemental Material.
In approximately 70\% of the realizations, DWs of both
polarities displaced by more than $20\,\mathrm{nm}$.
Separately, the absolute displacement difference between
the two polarities exceeded $20\,\mathrm{nm}$ in
approximately 30\% of the realizations.
The favored polarity varies between disorder realizations,
indicating that the selection depends on the local
anisotropy geometry.

To identify the origin of polarity-selective propagation,
we consider a simplified model with a step-like PMA profile
\cite{Franken_2011, Moretti_2017, Giuliano_2023,  Joseph_2026}.
Unlike in the axis-dependent profiles considered above, the PMA boundary can be tilted
away from the transverse direction.
As shown in Fig.~\ref{fig:LLG_dynamics-2}(a), the boundary is defined by $x+y\tan\chi_0=0$, where $\chi_0$
is the signed angle of its normal relative to the $x$ axis.
Denoting by $\kappa$ the anisotropy difference between
the two regions, we write
\begin{equation}
K_{\rm u}(x,y)
=K_0+\frac{\kappa}{2}
\left[
1+\tanh\left(
\frac{x+y\tan\chi_0}{\epsilon}
\right)
\right],
\end{equation}
where $K_0$ is the PMA constant on the
$x+y\tan\chi_0<0$ side in the sharp-boundary limit
$\epsilon\to0$.

Figures~\ref{fig:LLG_dynamics-2}(b) and (c) show the
micromagnetic DW dynamics for boundaries inclined at
$\chi_0=\pm15^\circ$.
Far from the PMA boundary, the DW dynamics approach those
in a uniform system.
In this region, the $\odot\otimes$ wall develops
$\chi>0$, whereas the $\otimes\odot$ wall develops
$\chi<0$
\cite{Boulle_2013, Martinez_2014, Baumgartner_2018, Volkov_2021,Kim_2022}.
Near the boundary, however, the two polarities exhibit qualitatively different behavior.
For $\chi_0=+15^\circ$, the $\odot\otimes$ wall is pinned,
whereas the $\otimes\odot$ wall crosses the boundary.
Reversing the sign of $\chi_0$ interchanges the pinned
and propagating polarities.
These results suggest stronger pinning when the
current-induced DW tilt and the boundary inclination
have the same sign.

To elucidate the underlying mechanism, we analyze the step-like PMA system within the 1DM framework
\cite{Malozemoff_1979, Tatara_2008, Shibata_2011, Thiaville_2012, Boulle_2013, Martinez_2014, Risinggard_2017, Volkov_2021, Umetsu_2025_JMMM, Umetsu_2025_PRB}.
The model describes the DW by its magnetization angle
$\phi$, position $X$, and tilt angle $\chi$
[Fig.~\ref{fig:LLG_dynamics-2}(a)]
\cite{Boulle_2013, Martinez_2014, Volkov_2021}.
We incorporate STT and the two-dimensional PMA profile
into the tilted-wall collective-coordinate framework.
The equations of motion are derived in
Sec.~\ref{sec:1DM} of the Supplemental Material:
\begin{equation}
\begin{split}\dot{\phi} & =\frac{\gamma \mu_0}{1+\alpha^{2}}\left(
-\alpha\cos\chi\mathcal{A}
-q\mathcal{B}
\right),\\
\dot{\zeta} & =\frac{\gamma \mu_0}{1+\alpha^{2}}\left(
q\mathcal{A}-\frac{\alpha}{\cos\chi}\mathcal{B}\right),\\
\dot{\chi} & =-\frac{\gamma \mu_0}{\alpha}\frac{12\cos^{3}\chi}{\pi^{2}\sin^{2}\chi+l_{y}^{2}}\frac{\partial\bar{E}}{\partial\chi},
\end{split}
\label{eq:1DM}
\end{equation}
where $\zeta=X/\lambda$ is the normalized DW position, $\lambda$ is the DW half width, and
$q=\pm1$ denotes the DW polarity ($q=+1$ for $\odot\otimes$ and $q=-1$ for $\otimes\odot$).
Here,  $\bar{E}\left(\phi,\zeta,\chi\right)= E/\left(2 \mu_0M_{{\rm s}}S_{\rm c}\lambda\right)$ is the magnetic energy expressed in field units, $S_{\rm c}= L_{y}L_{z}$ is the cross-sectional area, and $l_y=L_y/\lambda$.
We further define
$\mathcal{A}=\partial\bar{E}/\partial\phi+q \xi_{{\rm st}} j$ and
$\mathcal{B}=\partial\bar{E}/\partial\zeta+q(\pi /2) \xi_{{\rm so}} j \cos\phi$, where
$\xi_{{\rm st}}=-\hbar P/\left(2\mu_0 M_{{\rm s}}\lambda e \right)$ is the STT efficiency.
Although the local anisotropy can modify $\lambda$, we treat
it as constant and neglect the resulting width variations.
Equation~(\ref{eq:1DM}) is invariant under $q\to-q$, $\phi\to \pi -\phi$, $\chi \to -\chi$, and $\chi_0\to -\chi_0$,
which explains the symmetry observed in Fig.~\ref{fig:LLG_dynamics-2}(b) and (c).

Up to an additive constant, the effective
perpendicular-anisotropy contribution is
\begin{equation}
E_{\rm k}
=\frac{(2K_{0,\rm eff}+\kappa)S_{\rm c}\lambda}{\cos\chi}
-\frac{\kappa\lambda^2L_zF(\zeta,\chi)}{\cos^2\chi},
\label{eq:Ek}
\end{equation}
where $K_{0,\rm eff}=K_0-\mu_0M_{\rm s}^2/2$ and
\begin{equation}
F(\zeta,\chi)
=-\frac{2\operatorname{artanh}
\left[\tanh(\zeta\cos\chi)\tanh B\right]}
{\tan\chi-\tan\chi_0},
\label{eq:F}
\end{equation}
with $B=(l_y/2)\cos\chi(\tan\chi-\tan\chi_0)$.
The expression at $\chi=\chi_0$ is understood by
continuity.
The second term in $E_{\rm k}$ describes the
position-dependent anisotropy contribution responsible
for pinning.
At $\zeta=0$, the magnitude of the pinning force is
maximal when $\chi=\chi_0$, where the DW is parallel
to the anisotropy boundary.

The total energy also includes exchange, a
demagnetization-induced hard-axis anisotropy, and DMI;
their explicit forms are given in
Sec.~\ref{sec:1DM} of the Supplemental Material.
We denote the hard-axis anisotropy field by $H_{\rm a}$
and define the DMI field as
$H_{\rm D}=D/(\mu_0M_{\rm s}\lambda)$
\cite{Boulle_2013, Martinez_2014, Risinggard_2017,
Umetsu_2025_JMMM, Umetsu_2025_PRB}.
Using the method of DeJong et al. \cite{DeJong_2015},
we obtain $H_{\rm a}=0$ and $\lambda=6.1\,\mathrm{nm}$
for the present model.
For the static analysis below, we define
$H_{\rm c}=(K_{0,\rm eff}+\kappa/2+A/\lambda^2)/
(\mu_0M_{\rm s})$ and
$H_\kappa=\kappa/(2\mu_0M_{\rm s}l_y)$.

\begin{figure}[htbp]
\centering{}\includegraphics[width=0.8\linewidth]{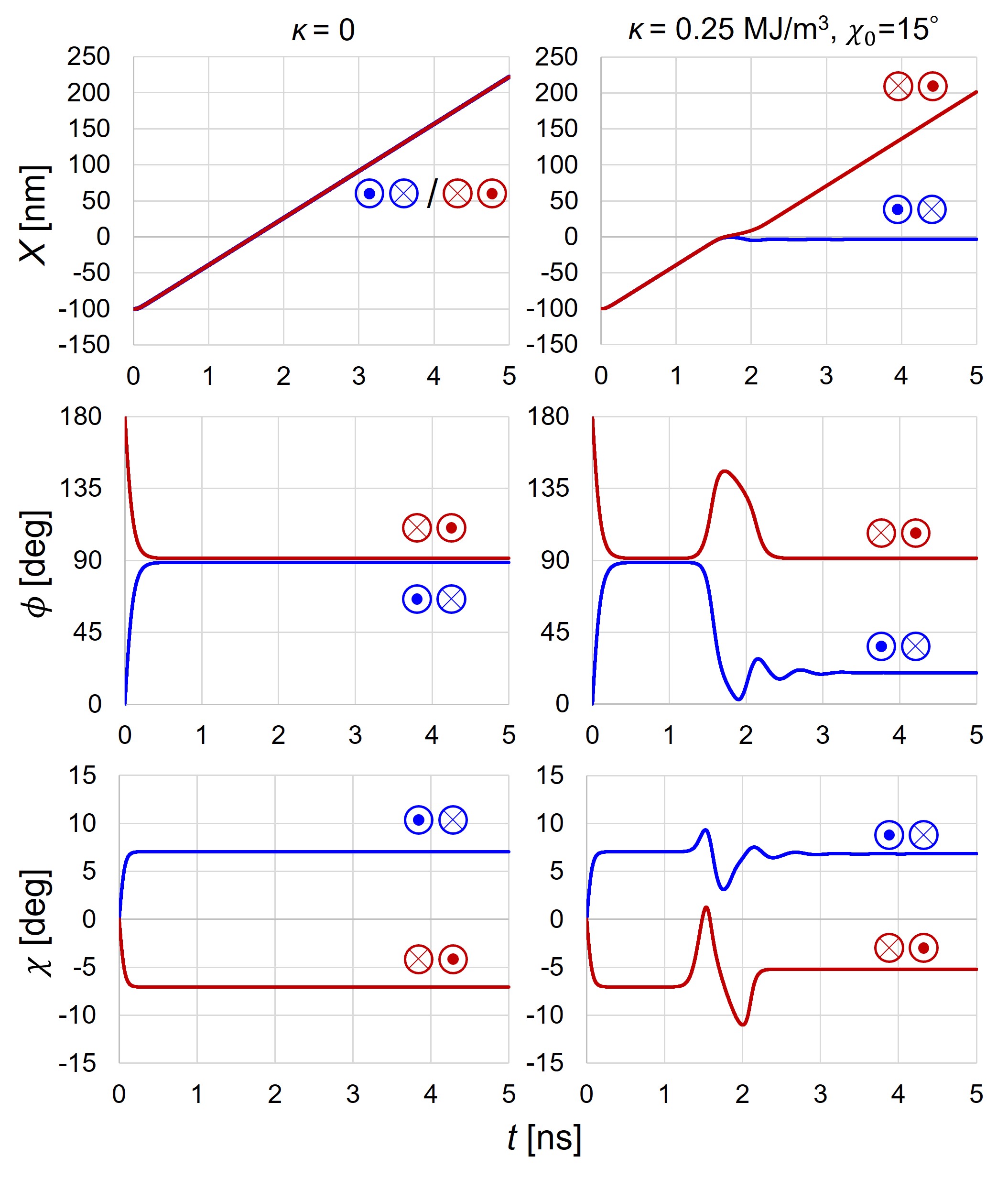}\caption{
DW dynamics computed from the 1DM  at $j=70\times10^{10}\,\mathrm{A/m^2}$.
The origin $X=0$ is at the center of the $K_{{\rm u}}$ boundary.
The left panels correspond to $\kappa=0$,
while the right panels correspond to $\kappa=0.25\,$MJ/m$^{3}$ with $\chi_{0}=15^{\circ}$.}
\label{fig:dynamics_1DM}
\end{figure}

Figure~\ref{fig:dynamics_1DM} shows the 1DM dynamics of the
$\odot\otimes$ and $\otimes\odot$ walls at $j=70\times10^{10}\,\mathrm{A/m^2}$
for $\kappa=0$ and $\kappa=0.25\,\mathrm{MJ/m^3}$ with $\chi_{0}=15^{\circ}$.
For $\kappa=0$, both walls reach steady motion within
$0.5\,\mathrm{ns}$, with constant velocity, $\phi$, and $\chi$.
In the uniform system, the two DW polarities have identical
steady-state velocities, while their magnetization angles
are symmetric about $\phi=\pi/2$ and their tilt angles
have opposite signs \cite{Martinez_2014}.

For $\kappa=0.25\,\mathrm{MJ/m^3}$ and $\chi_0=15^\circ$,
the initial dynamics are similar to those in the uniform
system because the DW is far from the boundary.
As the walls approach it, both tilt angles shift toward
$\chi_0>0$ during
$1.2\,\mathrm{ns}\lesssim t\lesssim1.5\,\mathrm{ns}$.
The anisotropy contribution favors alignment with the
boundary, whereas the SOT-driven dynamics coupled to DMI
favor $\chi<0$ for the $\otimes\odot$ wall.
This competition keeps that wall less aligned with the
boundary, weakening its pinning and allowing it to cross.

\begin{figure}[htbp]
\centering{}\includegraphics[width=1\linewidth]{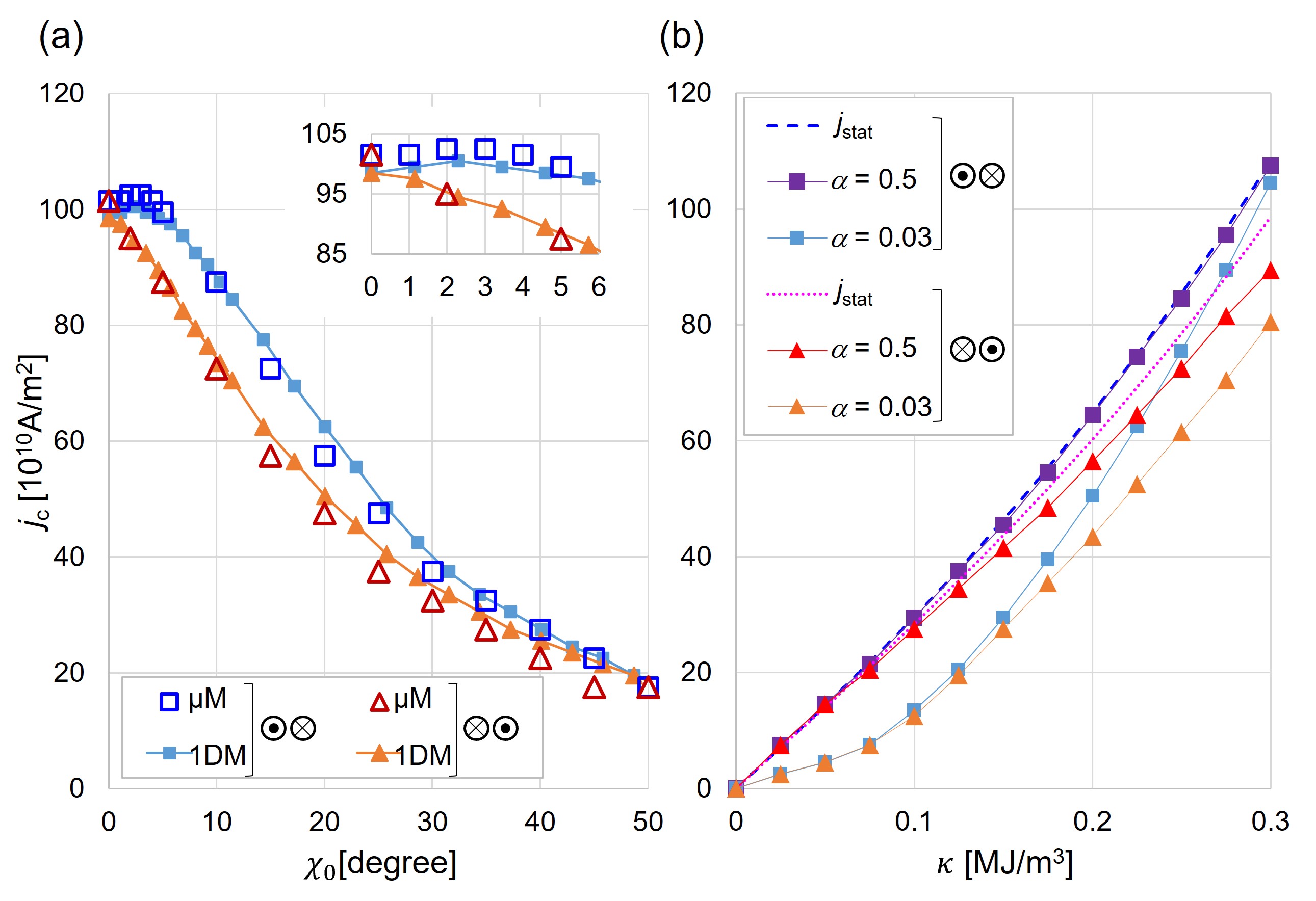}
\caption{
Propagation threshold $j_{\rm c}$ as a function of  (a) $\chi_0$ for
$\kappa=0.25\,\mathrm{MJ/m^3}$ and (b) $\kappa$ for
$\chi_0=15^\circ$.
Panel (a) uses $\alpha=0.03$, while panel (b) shows
1DM results for $\alpha=0.03$ and $0.5$.
The pulse duration is $10\,\mathrm{ns}$.
Open and filled markers represent micromagnetic and
1DM results, respectively.
In (a), the inset shows an enlarged view near $\chi_0=0$.
In (b), the dashed and dotted lines indicate
$j_{\rm stat}$.
}
\label{fig:jc-chi0}
\end{figure}

We define the critical current density for DW propagation,
$j_{\rm c}$, as the smallest positive current density
for which a DW initially at $X=-100\,\mathrm{nm}$ has $X>0$ at the end of a pulse.
Unless otherwise stated, the pulse duration is
$10\,\mathrm{ns}$.
Figure~\ref{fig:jc-chi0}(a) shows $j_{{\rm c}}$ as a function of  $\chi_{0}$ for $\kappa=0.25\,\mathrm{MJ/m^3}$.
The 1DM results are in good agreement with the micromagnetic simulations
and reveal a clear difference in $j_{{\rm c}}$ between the $\odot\otimes$
and $\otimes\odot$ walls for $0^{\circ}<\chi_{0}\lesssim50^{\circ}$.
This difference reflects the polarity-dependent
alignment of the DW with the anisotropy boundary:
at $\zeta=0$, the pinning force associated with the
second term in $E_{\rm k}$ is strongest when
$\chi=\chi_0$. 

For the $\otimes\odot$ wall, which has $\chi<0$ in a system with uniform $K_\mathrm{u}$,
$j_{{\rm c}}$ decreases monotonically with increasing $\chi_{0}$.
By contrast, $j_{\rm c}$ for the $\odot\otimes$ wall
exhibits a maximum near $\chi_0\simeq2^\circ$.
This behavior reflects the changing alignment between
the wall with $\chi>0$ and the anisotropy boundary.
For sufficiently large $\chi_0$, pinning weakens and the
polarity splitting becomes negligible.

Figure~\ref{fig:jc-chi0}(b) shows the $\kappa$ dependence of $j_{\mathrm c}$
for both DW polarities, obtained from the 1DM for $\alpha=0.03$ and $0.5$.
As $\kappa$ increases, $j_{\mathrm c}$ increases for both polarities,
and the splitting between the two thresholds becomes more pronounced.
The lower $j_{\rm c}$ at smaller $\alpha$ indicates that
the propagation threshold depends on both the static
pinning landscape and transient DW dynamics.
A higher SOT-driven DW velocity at lower damping
\cite{Thiaville_2012, Boulle_2013} can allow the wall to
cross the PMA boundary before reaching a pinned state.
The coupled transient dynamics of $\phi$ and $X$ may
also persist longer at lower damping and facilitate
boundary crossing \cite{Moretti_2017}.
Deriving a general analytical criterion for propagation
would require a stability analysis and a nonlinear
treatment of the transient dynamics, including their
dependence on the initial conditions.
Such an analysis is beyond the scope of this work.

To separate static and dynamical effects, we define
$j_{\rm stat}$ as the upper current limit of the
stationary pinned branch considered here.
For $H_{\rm a}=0$, expanding $j_{\rm stat}$ in
$k=H_\kappa/H_{\rm c}$ gives
$j_{\rm stat}=a_1k+a_2k^2+O(k^3)$, where
\begin{equation}
a_1=\frac{l_yR}{|\beta|},\qquad
a_2=-q\frac{\eta l_y^2RU}{(1-d)\beta^2}.
\label{eq:coefficients}
\end{equation}
Here, $\eta=\xi_{\rm st}/H_{\rm c}$,
$\beta=\pi\xi_{\rm so}/(2H_{\rm c})$,
$d=\pi H_{\rm D}/(2H_{\rm c})$,
$a=-(l_y/2)\tan\chi_0$,
$R=r(a)$, $U=(l_y/2)r'(a)$, and
$r(a)=\tanh(a)/a$, with $r(0)=1$.
Primes denote derivatives with respect to the argument.
The expansion is obtained for $\beta<0$ and $0<d<1$;
its derivation and range of validity are discussed in
Sec.~\ref{sec:j_static} of the Supplemental Material.
The quadratic coefficient changes sign with the DW
polarity and is proportional to the STT efficiency.
The $j_{\rm stat}$ curves in Fig.~\ref{fig:jc-chi0}(b)
include the cubic term given in
Sec.~\ref{sec:j_static} of the Supplemental Material.
Because the stationary-state conditions do not involve
Gilbert damping, $j_{\rm stat}$ is independent of $\alpha$.

As shown in Fig.~\ref{fig:jc-chi0}(b), $j_{\rm c}$
agrees well with $j_{\rm stat}$ for $\alpha=0.5$
in the small-$\kappa$ regime.
The polarity splitting in $j_{\rm stat}$ first appears
at order $k^2$.
For $|\chi_0|\ll1$,
$a_1=(l_y/|\beta|)(1-l_y^2\chi_0^2/12)+O(\chi_0^4)$ and
$a_2=-q\eta l_y^4\chi_0/[6(1-d)\beta^2]+O(\chi_0^3)$.
Thus, the leading angle-dependent correction common
to both polarities is of order $\chi_0^2k$, whereas
the polarity-dependent contribution begins at order
$\chi_0k^2$.
For $\xi_{\rm st}<0$, $\chi_0>0$, and $0<d<1$,
$a_2>0$ for $q=+1$ and $a_2<0$ for $q=-1$,
giving a higher static threshold for the
$\odot\otimes$ wall.

The correspondence between $j_{\rm c}$ and
$j_{\rm stat}$ is also confirmed for variations in
$\Theta_{\rm SH}$, $P$, and $D$ at $\alpha=1$ using
$50\,\mathrm{ns}$ pulses
[Fig.~\ref{fig:jc_1DM}(a)--(c) of the Supplemental
Material].
In the weak-damping regime, the polarity splitting in
$j_{\rm c}$ need not follow the static prediction.
In particular, the splitting remains finite at $P=0$,
where STT is absent
[Fig.~\ref{fig:jc_1DM}(b) of the Supplemental Material].
Thus, STT is not required for polarity-selective
propagation in the weak-damping case, although it
contributes to the static threshold splitting.
Another difference concerns the DMI dependence:
the $O(k^2)$ coefficient suggests an increasing
splitting with DMI, whereas the numerical 1DM results
show a decrease
[Fig.~\ref{fig:jc_1DM}(c) of the Supplemental Material].
As $d$ increases, higher-order terms can become important,
potentially reducing the accuracy of the static expansion
truncated at $O(k^2)$.
In addition, transient dynamics affect the propagation
threshold.
Stronger DMI stabilizes the N\'eel configuration and
can increase the DW velocity, thereby facilitating
dynamical passage across the boundary
\cite{Thiaville_2012, Boulle_2013, Martinez_2014,
Risinggard_2017, Umetsu_2025_JMMM, Umetsu_2025_PRB}.
A similar reduction in polarity splitting occurs as
$H_{\rm a}$ decreases
[Fig.~\ref{fig:jc_1DM}(d) of the Supplemental Material],
consistent with this interpretation.

We have shown that anisotropy geometry enables
polarity-selective SOT-driven DW propagation.
Opposite current-induced tilts produce different
pinning strengths at the same boundary, and reversing
its inclination reverses the selection.
The collective-coordinate model captures this mechanism
and distinguishes static pinning limits from propagation
thresholds.
These findings motivate comparisons of opposite-polarity
walls at the same pinning site and suggest a route to
polarity-selective control through anisotropy design.

\bibliography{reference}

\end{document}




\title{Supplemental Material for
``Polarity-selective domain wall propagation driven by
spin-orbit torque''}

\maketitle

\section{Construction of spatially correlated PMA distributions}
\label{sec:PMA}

We generate spatially correlated perpendicular magnetic
anisotropy (PMA) distributions $K_{\rm u}$ for micromagnetic
simulations, parameterized by a mean $\bar K$, a nominal
fluctuation amplitude $\sigma_{\rm k}$, and a correlation
length parameter $l_{\rm k}$.

We first set $K_{\rm u}=\bar K$ in every computational cell
and add independent Gaussian random variables with zero
mean and standard deviation $c$.
The resulting distribution is evolved from $t=0$ to $t=1$
using the diffusion equation, where $t$ is a dimensionless
auxiliary time.
The diffusion coefficient sets the correlation length,
while $c$ controls the fluctuation amplitude.
Below, we derive the corresponding normalization.

Let $A_m(\bm{x},t)$ denote a field at position $\bm{x}$
and time $t$ in realization $m$.
We define the ensemble average by
\begin{equation}
\left\langle A\left(\bm{x},t\right)\right\rangle =\lim_{M\to\infty}\frac{1}{M}\sum_{m=1}^{M}A_{m}\left(\bm{x},t\right),
\end{equation}
and the spatial average within realization $m$ by
\begin{equation}
\overline{A_{m}\left(t\right)}=\frac{1}{S_{\rm FM}} \int{\rm d}S\,A_{m}\left(\bm{x},t\right),
\end{equation}
where $S_{\rm FM}$ is the in-plane area of the ferromagnetic
layer.
Their combination gives
\begin{equation}
\left\langle \overline{A\left(t\right)}\right\rangle =\lim_{M\to\infty}\frac{1}{M}\sum_{m=1}^{M}
\frac{1}{S_{\rm FM}}
\int{\rm d}S \, A_{m}\left(\bm{x},t\right).
\end{equation}

At $t=0$, an independent Gaussian random variable
$c_m(\bm{x}_i)$ is assigned to each cell center
$\bm{x}_i$ ($i=1,\ldots,N$), with
\begin{equation}
\left\langle c_{m}\left(\bm{x}_{i}\right)\right\rangle =0,\quad\left\langle c_{m}\left(\bm{x}_{i}\right)c_{m}\left(\bm{x}_{j}\right)\right\rangle =c^{2}\delta_{i,j}.
\end{equation}

The anisotropy fluctuation is
$\Psi_m(\bm{x},t)=K_{{\rm u},m}(\bm{x},t)-\bar K$.
For the analytical normalization, we use the continuum
approximation for a statistically homogeneous noise field
on an unbounded two-dimensional domain, neglecting
finite-domain boundary effects.
In a delta-function representation of the cell noise,
the initial fluctuation field is
\begin{equation}
\Psi_{m}\left(\bm{x},0\right)=\Omega\sum_{i=1}^{N}c_{m}\left(\bm{x}_{i}\right)
\delta^{(2)}(\bm{x}-\bm{x}_i),\label{eq:Phi_m}
\end{equation}
where $\Omega$ is the in-plane area of each computational
cell and $S_{\rm FM}=N\Omega$.

We define the ensemble-averaged spatial autocorrelation
function as
\begin{equation}
R\left(\bm{x},t\right)\!=\!\left\langle \frac{1}{S_{\rm FM}}\int\!{\rm d}S'\,\Psi_{m}\left(\bm{x}',t\right)\Psi_{m}\left(\bm{x}+\bm{x}',t\right)\right\rangle .
\end{equation}
At $t=0$, this reduces to
$R(\bm{x},0)=\Omega c^2\delta^{(2)}(\bm{x})$.
We evolve $\Psi_m(\bm{x},t)$ according to the diffusion
equation
\begin{equation}
\frac{\partial}{\partial t}\Psi_{m}\left(\bm{x},t\right)=D_{\rm diff}\nabla^{2}\Psi_{m}\left(\bm{x},t\right).
\end{equation}
The corresponding solution is
\begin{equation}
\Psi_{m}\left(\bm{x},t\right)=\sum_{i=1}^{N}\frac{\Omega c_{m}\left(\bm{x}_{i}\right)}{4\pi D_{\rm diff}t}\exp\left[-\frac{\left|\bm{x}-\bm{x}_{i}\right|^{2}}{4D_{\rm diff}t}\right].
\end{equation}

At $t=1$, the ensemble-averaged spatial mean remains zero,
$\langle\overline{\Psi(1)}\rangle=0$, and the autocorrelation
function is
\begin{equation}
R(\bm{x},1)
=\frac{c^2\Omega}{8\pi D_{\rm diff}}
\exp\left[-\frac{|\bm{x}|^2}{8D_{\rm diff}}\right].
\end{equation}
We define the correlation length $l_{\rm k}$ by
$R(\bm{x},1)/R(\bm{0},1)
=\exp[-|\bm{x}|^2/(4l_{\rm k}^2)]$,
which gives $D_{\rm diff}=l_{\rm k}^2/2$.
Within the infinite-domain approximation, the nominal
fluctuation amplitude is defined by
$\sigma_{\rm k}^2=\langle\overline{\Psi^2(1)}\rangle
=R(\bm{0},1)$.
The required normalization is therefore
\begin{equation}
c=2\sqrt{\frac{\pi}{\Omega}}\,l_{\rm k}\sigma_{\rm k}.
\end{equation}

The diffusion equation is solved on a finite domain with
zero normal derivative at the boundaries, using the
infinite-domain expression for $c$ without boundary
corrections.

\section{Domain wall displacement statistics in random anisotropy landscapes}

\begin{figure}[htbp]
\centering{}\includegraphics[width=1\linewidth]{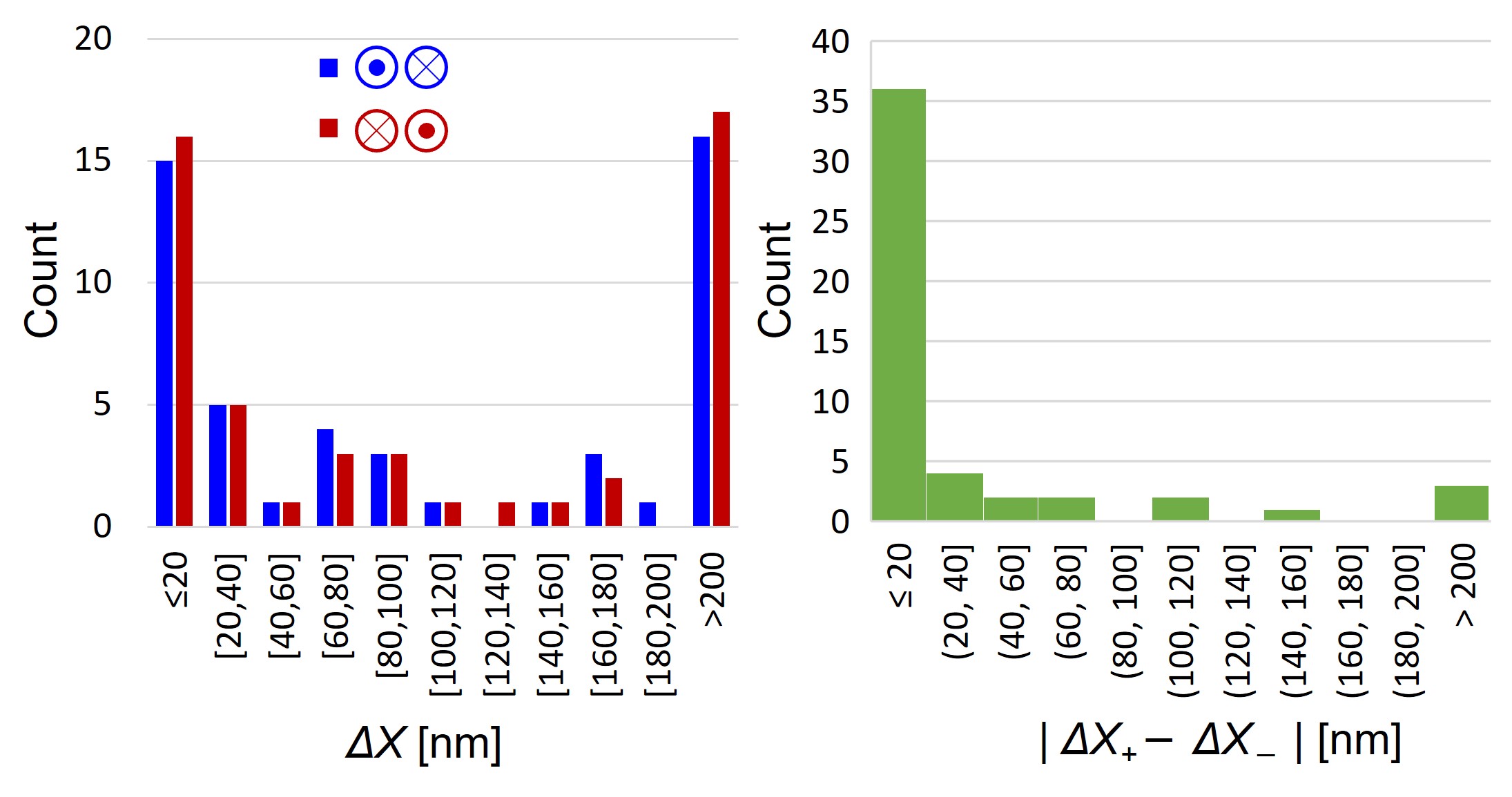}
\caption{
(a) Histograms of the DW displacement after a
$10\,\mathrm{ns}$ current pulse for the $\odot\otimes$
and $\otimes\odot$ polarities.
(b) Histogram of the absolute displacement difference
$|\Delta X_+-\Delta X_-|$ within each disorder realization.
The simulations use $j=45\times10^{10}\,\mathrm{A/m^2}$,
$\bar K=1.06\,\mathrm{MJ/m^3}$,
$\sigma_{\rm k}=0.2\,\mathrm{MJ/m^3}$, and
$l_{\rm k}=20\,\mathrm{nm}$.
}
\label{fig:histogram}
\end{figure}

Using the diffusion-based procedure described above,
we generated 50 random anisotropy landscapes.
For each realization, we simulated the current-driven
motion of up--down ($\odot\otimes$) and down--up
($\otimes\odot$) domain walls (DWs) separately
and recorded their displacements after a $10\,\mathrm{ns}$
current pulse.
Figure~\ref{fig:histogram}(a) shows the displacement
histograms for the two polarities, while panel (b) shows
the histogram of $|\Delta X_+-\Delta X_-|$.
Here, $\Delta X_+$ and $\Delta X_-$ denote the displacements
of the $\odot\otimes$ and $\otimes\odot$ DWs, respectively.

\section{Collective-coordinate formulation of the
one-dimensional model}
\label{sec:1DM}

The extended one-dimensional model (1DM) describes
the DW using three collective coordinates.
Using the generalized Euler--Lagrange formalism
\cite{Tatara_2008,Shibata_2011,Volkov_2021}, we obtain
the equations of motion for the collective coordinates
$(Q_1,Q_2,Q_3)=(\phi,\zeta,\chi)$:
\begin{equation}
\frac{\partial L_{\rm B}}{\partial Q_i}
-\frac{{\rm d}}{{\rm d}t}
\frac{\partial L_{\rm B}}{\partial\dot Q_i}
-\frac{\partial D_{\rm R}}{\partial\dot Q_i}
=\frac{\partial E}{\partial Q_i}.
\label{eq:1DM-0}
\end{equation}
Here, $E=\int\mathcal E\,{\rm d}V$ is the magnetic energy,
$D_{\rm R}=\int\mathcal D_{\rm R}\,{\rm d}V$ is the
generalized Rayleigh function, and $L_{\rm B}$ is the
Berry-phase contribution.
The volume integral is
$\int{\rm d}V=L_z\int_{-\infty}^{\infty}{\rm d}x
\int_{-L_y/2}^{L_y/2}{\rm d}y$.

We parameterize the magnetization as
$\bm m=(\sin\theta\cos\phi,\sin\theta\sin\phi,\cos\theta)$.
The azimuthal angle $\phi$ is taken to be spatially uniform,
while the polar angle follows the domain wall ansatz
\begin{equation}
\theta=2\tan^{-1}
\exp\left\{
q\left[
\left(\frac{x}{\lambda}-\zeta\right)\cos\chi
+\frac{y}{\lambda}\sin\chi
\right]
\right\}.
\end{equation}
Here, $\zeta=X/\lambda$ is the normalized DW position,
with $X$ denoting the wall position at $y=0$.
The DW tilt angle $\chi$ is measured from the $x$ axis
to the wall normal, and the DW half width $\lambda$
is treated as constant.
The polarity is $q=+1$ for an $\odot\otimes$ wall and
$q=-1$ for an $\otimes\odot$ wall.
For this profile, the Berry-phase contribution is
\begin{equation}
L_{\rm B}
=-\frac{2qM_{\rm s}S_{\rm c}\lambda}{\gamma}
\zeta\dot\phi,
\label{eq:Berry}
\end{equation}
where $S_{\rm c}=L_yL_z$.

The collective-coordinate equations derived here assume
$\beta_{\rm so}=0$, as in the main-text results.
Gilbert damping, spin-transfer torque (STT), and
damping-like spin-orbit torque (SOT) are incorporated
through the generalized Rayleigh function
\cite{Boulle_2013}:
\begin{equation}
\begin{aligned}
\mathcal D_{\rm R}
=\frac{\alpha M_{\rm s}}{2\gamma}
\Biggl(
&\dot{\bm m}
+\frac{\gamma\mu_0\xi_{\rm st}j\lambda}{\alpha}
\bm m\times\frac{\partial\bm m}{\partial x}\\
&+\frac{\gamma\mu_0\xi_{\rm so}j}{\alpha}
\bm m\times\bm e_y
\Biggr)^2.
\end{aligned}
\label{eq:R}
\end{equation}
Substituting the domain wall ansatz into Eq.~(\ref{eq:R})
and integrating over the volume gives the
velocity-dependent part of $D_{\rm R}$:
\begin{equation}
\begin{aligned}
\frac{\gamma D_{\dot{\bm m}}}
{M_{\rm s}S_{\rm c}\lambda}
={}&
\frac{\alpha}{\cos\chi}
\Biggl[
\dot\zeta^2\cos^2\chi+\dot\phi^2\\
&+\frac{\dot\chi^2}{12}
\left(
\pi^2\tan^2\chi+\frac{l_y^2}{\cos^2\chi}
\right)
\Biggr]\\
&+2q\gamma\mu_0\xi_{\rm st}j\dot\phi\\
&+\pi q\gamma\mu_0\xi_{\rm so}j
\dot\zeta\cos\phi,
\end{aligned}
\label{eq:Dmdot}
\end{equation}
where $l_y=L_y/\lambda$.
Terms independent of the collective-coordinate velocities
have been omitted.

In addition to the effective perpendicular-anisotropy
energy $E_{\rm k}$ given in Eq.~(\ref{eq:Ek}) of the
main text, the total magnetic energy $E$ includes
\begin{align}
E_{\rm exc}
&=\frac{2AS_{\rm c}}{\lambda\cos\chi},\\
E_{\rm a}
&=\frac{\mu_0M_{\rm s}S_{\rm c}\lambda H_{\rm a}}
{\cos\chi}\cos^2(\phi-\chi),\\
E_{\rm DMI}
&=-\frac{q\pi\mu_0M_{\rm s}S_{\rm c}\lambda H_{\rm D}}
{\cos\chi}\cos(\phi-\chi).
\end{align}
These are the exchange, demagnetization-induced
hard-axis anisotropy, and Dzyaloshinskii--Moriya
interaction (DMI) energies, respectively.
Here, $H_{\rm a}$ is the hard-axis anisotropy field
and $H_{\rm D}=D/(\mu_0M_{\rm s}\lambda)$ is the DMI
field
\cite{Boulle_2013, Martinez_2014, Risinggard_2017,
Umetsu_2025_JMMM, Umetsu_2025_PRB}.
Combining these contributions gives the normalized
energy $\bar E=E/(2\mu_0M_{\rm s}S_{\rm c}\lambda)$:
\begin{equation}
\bar E(\phi,\zeta,\chi)
=\frac{H_{\rm c}I(\phi,\chi)}{\cos\chi}
-\frac{H_\kappa F(\zeta,\chi)}{\cos^2\chi}
+\mathrm{const.},
\end{equation}
where $K_{0,\rm eff}=K_0-\mu_0M_{\rm s}^2/2$,
$H_{\rm c}=(K_{0,\rm eff}+\kappa/2+A/\lambda^2)/
(\mu_0M_{\rm s})$,
$H_\kappa=\kappa/(2\mu_0M_{\rm s}l_y)$, and
\begin{equation}
\begin{aligned}
I(\phi,\chi)
={}&1+\frac{H_{\rm a}}{2H_{\rm c}}
\cos^2(\phi-\chi)\\
&-\frac{q\pi H_{\rm D}}{2H_{\rm c}}
\cos(\phi-\chi).
\end{aligned}
\end{equation}
Using Eqs.~(\ref{eq:Berry}) and (\ref{eq:Dmdot}),
together with the magnetic energy specified above,
Eq.~(\ref{eq:1DM-0}) yields the equations of motion
in Eq.~(\ref{eq:1DM}) of the main text.

\section{Stationary pinning limit and its small-$k$ expansion}
\label{sec:j_static}

Setting $\dot\phi=\dot\zeta=\dot\chi=0$ in
Eq.~(\ref{eq:1DM}) of the main text gives the following
stationary-state conditions for $H_{\rm a}=0$:
\begin{subequations}
\begin{align}
0 & =\eta j+d\left(\sin\phi-\cos\phi\tan\chi\right),\label{eq:eq-1-1}\\
0 & =q\beta j\cos\phi-k\sec^{2}\chi F_{\zeta},\label{eq:eq-2-1}\\
0 & =\sin\chi-qd\sin\phi-k\left(2F\tan\chi+F_{\chi}\right),\label{eq:eq-3-1}
\end{align}
\end{subequations}
where subscripts denote partial derivatives,
$k=H_\kappa/H_{\rm c}$,
$\eta=\xi_{\mathrm{st}}/(H_{\mathrm c})$, $\beta=\pi \xi_{\mathrm{so}}/(2H_{\mathrm c})$,
and $d=\pi H_{\mathrm D}/(2H_{\mathrm c})$.
The function $F$ is defined in Eq.~(\ref{eq:F}) of the
main text, with
$B=(l_y/2)(\sin\chi-\cos\chi\tan\chi_0)$.
We assume $\beta<0$, $l_y>0$, and $0<d<1$, and consider
positive currents $j>0$.
In the limit $k\to0^+$ with all other parameters fixed,
we follow the stationary branch that approaches
$(\phi,\chi)=(0,0)$ for $q=+1$ and $(\pi,0)$ for $q=-1$.
We identify the static threshold with the maximum current
along this branch for $\zeta\leq0$.
Near $d=0$ or $d=1$, higher-order terms in the $k$ expansion
can become important even for small $k$.

We further define
$a=-(l_y/2)\tan\chi_0$, $r(a)=\tanh(a)/a$ with $r\left(0\right)=1$,
$R=r(a)$, $S=\sech^2 a$, $U=(l_y/2)r'(a)$, and
$V=(l_y^2/4)r''(a)-a r'(a)$.
Primes denote derivatives with respect to the argument.

At leading order, Eq.~(\ref{eq:eq-2-1}) gives
\begin{equation}
\frac{j}{k}=\frac{l_{y}R}{\left|\beta\right|\left(1+S\sinh^{2}\zeta\right)}+O\left(k\right).
\end{equation}
Let $\zeta_{\rm c}$ denote the position that maximizes
the current along the stationary branch.
Since $R,S>0$, the leading-order expression has a
nondegenerate maximum at $\zeta=0$.
A smooth $O(k)$ correction therefore gives
$\zeta_{\rm c}=O(k)$.

We introduce the small angular variables
\begin{equation}
h=q\chi,\qquad
\vartheta=
\begin{cases}
\phi, & q=+1\\
\pi-\phi, & q=-1
\end{cases}
,
\end{equation}
for which $j,h,\vartheta=O\left(k\right)$, $\sin\phi=\sin\vartheta$, and
$\cos\phi=q\cos\vartheta$.

Using $B=a+ql_{y}h/2-ah^{2}/2+O\left(h^{3}\right)$, we obtain
\begin{equation}
r\left(B\right)=R+qUh+\frac{1}{2}Vh^{2}+O\left(h^{3}\right).
\end{equation}
Expanding $F_\zeta$ in Eq.~(\ref{eq:eq-2-1}) gives the
current through cubic order in $k$:
\begin{equation}
\begin{split}j= & \frac{l_{y}k}{\left|\beta\right|}\Bigl[R+qUh+\frac{V}{2}h^{2}\\
 & \quad+\frac{R}{2}\vartheta^{2}-RS\zeta^{2}\Bigr]+O\left(k^{4}\right).
\end{split}
\label{eq:j-1}
\end{equation}
Similarly, expanding the anisotropy term gives
$2F\tan\chi+F_\chi=-l_yU\zeta+O(k^2)$.
Equations~(\ref{eq:eq-1-1}) and~(\ref{eq:eq-3-1}) then reduce
to 
\begin{align}
\vartheta
&=h-\frac{\eta}{d}j+O(k^3),
\label{eq:theta}\\
h
&=-\frac{\eta j+qkl_yU\zeta}{1-d}+O(k^3).
\label{eq:x}
\end{align}

Maximizing Eq.~(\ref{eq:j-1}) with respect to $\zeta$,
subject to Eqs.~(\ref{eq:theta}) and (\ref{eq:x}), gives
\begin{equation}
\zeta_{{\rm c}}=-\frac{l_{y}U^{2}}{2\left(1-d\right)RS}k+O\left(k^{2}\right).\label{eq:zeta_c}
\end{equation}
For $U\neq0$, $\zeta_{\rm c}<0$ for sufficiently small
$k>0$, so the maximizer lies strictly inside the region
$\zeta\leq0$.

Substituting Eq.~(\ref{eq:zeta_c}) into
Eqs.~(\ref{eq:j-1})--(\ref{eq:x}) and solving
order by order in $k$, we obtain
\begin{equation}
j_{\rm stat}
=a_1k+a_2k^2+a_3k^3+O(k^4),
\label{j_stat}
\end{equation}
with
\begin{subequations}
\begin{align}
a_{1} & =\frac{l_{y}R}{\left|\beta\right|},\\
a_{2} & =-q\frac{\eta l_{y}^{2}RU}{\left(1-d\right)\beta^{2}},\\
a_{3} & =\frac{\eta^{2}l_{y}^{3}}{\left(1-d\right)^{2}\left|\beta\right|^{3}}\left(RU^{2}+\frac{R^{2}V}{2}+\frac{R^{3}}{2d^{2}}\right)\nonumber \\
 & +\frac{l_{y}^{3}U^{4}}{4\left(1-d\right)^{2}\left|\beta\right|RS}. \label{coef_j}
\end{align}
\end{subequations}

\begin{figure}[htbp]
\centering{}\includegraphics[width=1\linewidth]{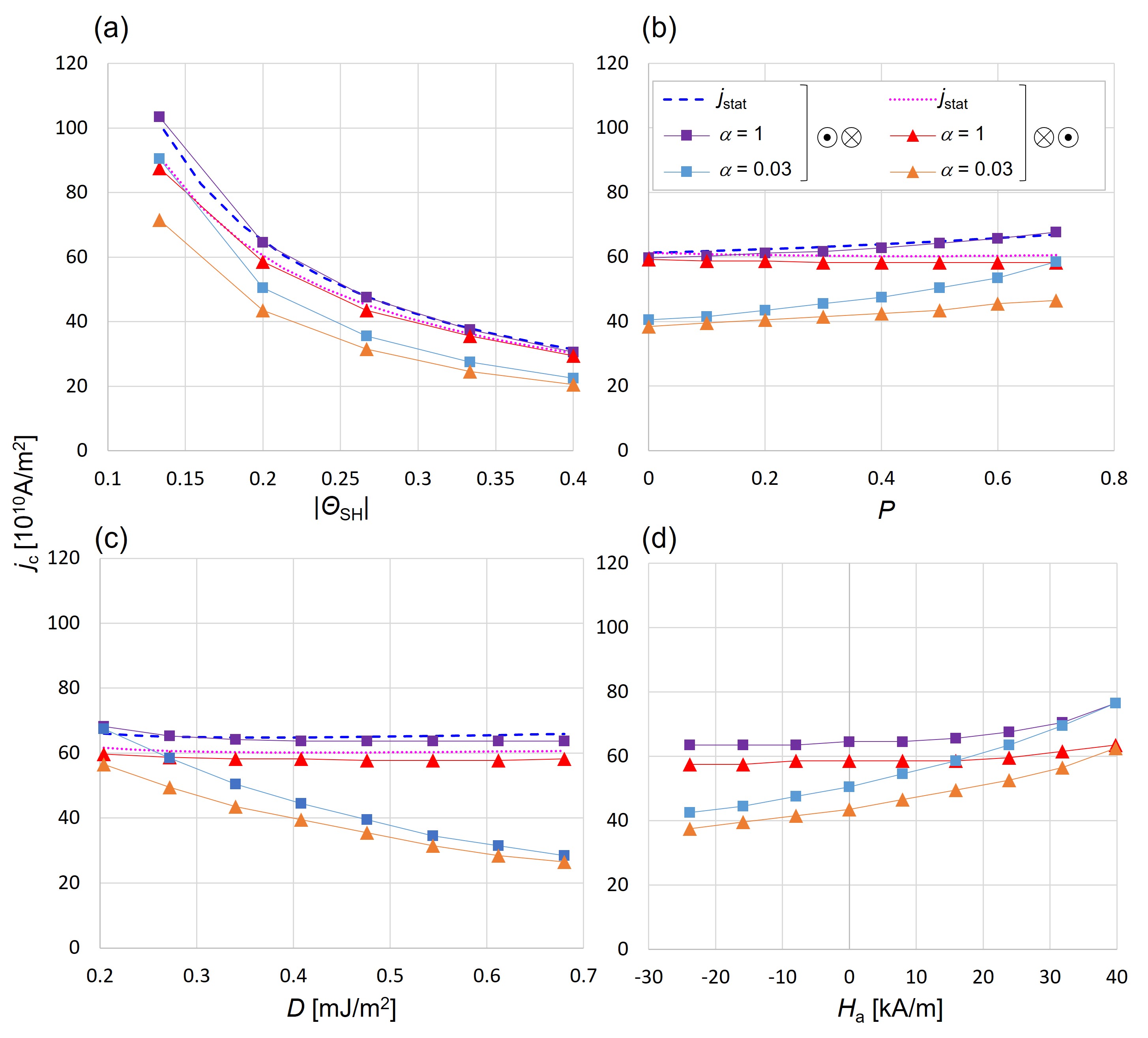}
\caption{
Critical current density for DW propagation calculated
using the 1DM as a function of
(a) the magnitude of the spin Hall angle $|\Theta_{\rm SH}|$,
(b) the spin polarization $P$,
(c) the DMI constant $D$, and
(d) the demagnetization-induced hard-axis anisotropy field
$H_{\rm a}$, for $\kappa=0.2\,\mathrm{MJ/m^3}$ and
$\chi_0=15^\circ$.
The pulse durations are $10\,\mathrm{ns}$ for $\alpha=0.03$
and $50\,\mathrm{ns}$ for $\alpha=1$.
The dashed and dotted lines indicate $j_{\rm stat}$.
All other parameters are those used in the main text;
$H_{\rm a}=0$ in panels (a)--(c).
}
\label{fig:jc_1DM}
\end{figure}

Figure~\ref{fig:jc_1DM} shows the propagation threshold
$j_{\rm c}$ obtained by numerical integration of the
1DM as a function of $|\Theta_{\rm SH}|$, $P$, $D$, and
$H_{\rm a}$.
As in the main text, propagation is identified by
$X>0$ at the end of the pulse for a DW initially at
$X=-100\,\mathrm{nm}$.
The pulse durations are $10\,\mathrm{ns}$ for
$\alpha=0.03$ and $50\,\mathrm{ns}$ for $\alpha=1$.
The longer pulse at $\alpha=1$ allows additional time
for the slower wall to reach the boundary.
Panels (a)--(c) also show $j_{\rm stat}$ from
Eq.~(\ref{j_stat}).
In panel (b), the propagation thresholds remain
different at $P=0$ for $\alpha=0.03$, showing that
STT is not required for polarity-selective propagation
in the weak-damping case.
No static-threshold curve is shown in panel (d), because
the expansion assumes $H_{\rm a}=0$.

\section{Effect of field-like torque on polarity-selective
domain wall propagation}
\label{sec:FLT}

Using the torque expression defined in the main text,
we evaluate the effect of field-like torque (FLT)
on the critical current density $j_{\rm c}$ for both
DW polarities by varying $\beta_{\rm so}$.

\begin{figure}[htbp]
\centering
\includegraphics[width=0.7\linewidth]{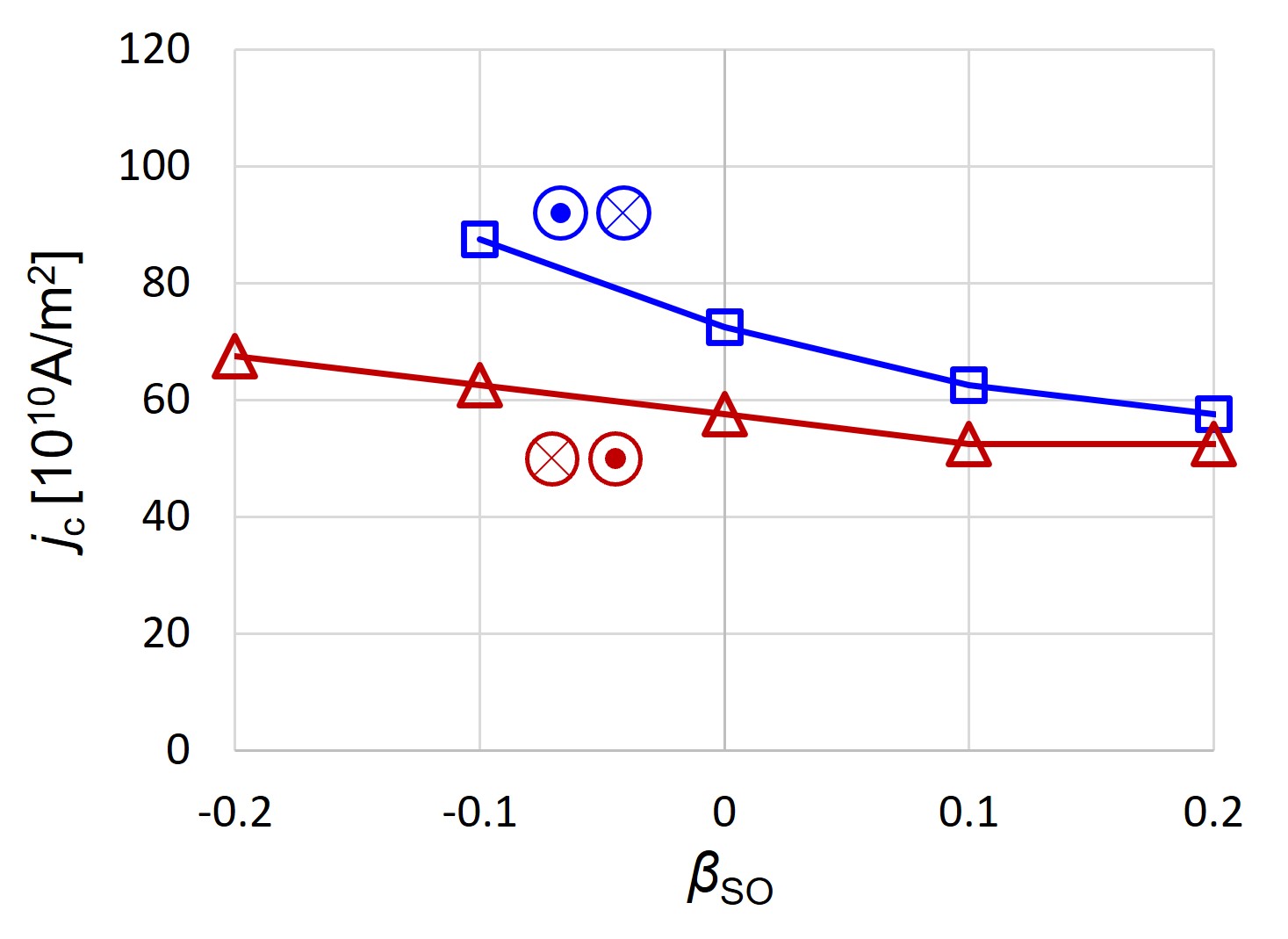}
\caption{
Critical current density for DW propagation obtained
from micromagnetic simulations as a function of
the signed ratio of the field-like to damping-like
effective-field amplitudes $\beta_{\rm so}$, for
$\kappa=0.2\,\mathrm{MJ/m^3}$ and $\chi_0=15^\circ$.
The pulse duration is $10\,\mathrm{ns}$, and
$\alpha=0.03$.
No data point is shown for the $\odot\otimes$ wall at
$\beta_{\rm so}=-0.2$, because no boundary crossing was
observed in the range
$0<j<150\times10^{10}\,\mathrm{A/m^2}$.
}
\label{fig:jc_FLT}
\end{figure}

Figure~\ref{fig:jc_FLT} shows $j_{\rm c}$ obtained
from micromagnetic simulations as a function of
$\beta_{\rm so}$.
Increasing $\beta_{\rm so}$ reduces the propagation
thresholds and their polarity splitting over the range where both thresholds could be determined.
For $\beta_{\rm so}=-0.2$, no boundary crossing was
observed for the $\odot\otimes$ wall within the
investigated range,
$0<j<150\times10^{10}\,\mathrm{A/m^2}$.
The corresponding threshold could not be determined
within this range, and no data point is shown.
At higher currents, STT can reduce the DW velocity in
the current direction \cite{Umetsu_2025_JMMM}, so increasing
the current further does not necessarily promote boundary
crossing.

The reduction in the propagation thresholds is consistent
with an FLT-induced increase in DW velocity.
Our previous study \cite{Umetsu_2025_JMMM} showed that,
for $\Theta_{\rm SH}<0$ and $\beta_{\rm so}>0$, the FLT
effective field along the negative $y$ direction makes
the current-driven DW configuration more N\'eel-like
and increases its velocity.
Faster motion can allow the DW to cross the PMA boundary
before relaxing to a pinned state, thereby lowering
the propagation threshold.

Although FLT can also modify the DW tilt \cite{Boulle_2013},
the change in DW tilt before reaching the boundary was
small in the comparison performed here.
At $j=85\times10^{10}\,\mathrm{A/m^2}$ and
$t=0.5\,\mathrm{ns}$, when the DWs had not yet reached
the boundary, the tilt angles for $\beta_{\rm so}=+0.2$
and $-0.2$ differed by less than $1^\circ$.
The small tilt difference in this comparison is
consistent with the velocity-based interpretation above.

\bibliography{reference}